\documentclass[twocolumn,showpacs,floats,letterpaper,floatfix,
groupedaddress,eqsecnum]{revtex4}
\usepackage{amssymb,amsmath}
\usepackage[dvips]{graphicx}
\usepackage{subfigure}

\usepackage{dcolumn,epsfig}

\begin{document}

\title{ Dispersionless motion in a driven periodic potential}

\author{S. Saikia$^1$ and Mangal C. Mahato}
\email{mangal@nehu.ac.in}
\affiliation{Department of Physics, North-Eastern Hill University, 
Shillong-793022, India\\
$^1$Department of Physics, St. Anthony's College, Shillong-793001, India.}

\begin{abstract}
Recently, dispersionless (coherent) motion of (noninteracting) massive 
Brownian particles, at intermediate time scales, was reported in a sinusoidal 
potential with a constant tilt. The coherent motion persists for a finite 
length of time before the motion becomes diffusive. We show that such coherent 
motion can be obtained repeatedly by applying an external zero-mean 
square-wave drive of appropriate period and amplitude, instead of a constant 
tilt. Thus, the 
cumulative duration of coherent motion of particles is prolonged. Moreover, 
by taking an appropriate combination of periods of the external field, one 
can postpone the beginning of the coherent motion and can even have 
coherent motion at a lower value of position dispersion than in the constant 
tilt case. 
\end{abstract}

\vspace{0.5cm}
\date{\today}

\maketitle

The inertial Brownian particle motion in periodic 
potentials\cite{risken, reim} has been an archetypal model to theoretically 
understand many phenomena in physical systems. The current-voltage 
characteristics of (RCSJ model of) Josephson junctions\cite{falco}, the 
electrical conductivity of superionic solids\cite{fulde}, 
motion of adatoms on the surface of a crystal\cite{lacasta}, etc., are some
of the important examples\cite{risken}. However, not all behaviour of the 
model particle motion in all time regimes are exhaustively investigated. 
A recent example being the discovery of dispersionless particle 
motion in a tilted periodic potential in the intermediate time regime 
by Lindenberg and coworkers\cite{linde}. During the coherent motion, the 
ensemble averaged position dispersion, $\Delta x(t)=<(x(t)-<x(t)>)^2>$, 
remains constant. 
 
This interesting phenomenon is shown (numerically) by particles 
moving on a cosinusoidal potential with a constant tilt (CT), $F_0$, 
in a medium with constant friction coefficient $\gamma_0$\cite{linde}
in a limited ($F_0,\gamma_0$) region. 
The particles, after crossing the immediate barrier move (after 
$t=\tau_1>\tau_K$, the Kramers mean passage time) coherently with 
velocity $v\approx\frac{F_0}{\gamma_0}$. The coherent motion 
continues until it is overwhelmed (at around $t=\tau_2$) by the diffusive 
motion of the particles. $\tau_1$ and $\tau_2$ are specified only 
as a rough guide\cite{linde}. In this work we investigate the effect 
of a zero-mean square-wave external drive (ZMSW) $F(t)$ of half period 
$\tau$ and amplitude $F_0$, instead of a CT. 

The coherent motion, naturally, gets interrupted upon 
reversal of direction of $F$ at $t=\tau$ ($\tau_1 <\tau<\tau_2$). 
Interestingly, as a main result of this work, the coherent motion once 
disturbed, by reversing the field at $t=\tau$, gets reestablished around 
$t=\tau+\tau_1$ in almost the same form as it was during $\tau_1<t<\tau_2$
in the CT case. And this loss and subsequent
recovery of coherent motion continues for a large number of
reversals of $F(t)$. The dispersion $\Delta x(t)$, however, increases 
rapidly during $\tau<t<\tau+\tau_1$. (During $0<t<\tau_1$, 
$\Delta x(t)\sim t^{\alpha}$, $\alpha\approx 2$).  

We consider the motion of an ensemble of Brownian particles each of mass 
$m$ moving in a potential
$V(x)=-V_0 \sin(kx)$ in a medium with friction coefficient\cite{wahn}
$\gamma(x)=\gamma_0(1-\lambda \sin(kx+\phi))$ at temperature $T$ (in units of
$k_B$) and subjected to an external force field $F(t)$. The
corresponding Langevin equation in dimensionless form is given 
by\cite{lambda}
\begin{equation}
\frac{d^2x}{dt^2}=-\gamma(x)\frac{dx}{dt}+\cos x +F(t)
+\sqrt{\gamma(x)T}\xi(t),
\end{equation}
with $\gamma(x)=\gamma_0(1-\lambda \sin(x+\phi))$.
Here all terms are written in units of $m$, $V_0$ and $k$, for example,
the scaled time $t$ is measured in units of
$\sqrt{\frac{m}{V_0k^2}}$, etc. The Gaussian distributed zero-mean 
random forces $\xi(t)$ satisfy $<\xi(t)\xi(t')>=2\delta(t-t')$. The applied 
square-wave force $F(t)$
is taken as $F(t)=\pm F_0$ for ($2n\tau\leq t<(2n+1)\tau$) and
$F(t)=\mp F_0$ for ($(2n+1)\tau \leq t<(2n+2)\tau$) with
$n=0, 1, 2, ... $. For constant applied force $F(t)=F_0$
(for all $t$) the equation is solved using the matrix continued
fraction method and also numerically, supporting each other
quantitatively\cite{risken, volmer, wanda, wanda1, shantu}. However,
for finite $\tau$, the equation could be solved only numerically. 
The integration of the equation, using the $4^{th}$ order Runge-Kutta
method\cite{nume}, was carried out in time steps of $\Delta t=.001$. 

We take $\lambda=0.9$. $\lambda\neq0$, however, is relevant only while
discussing ratchet current at the end. In fact, except for this minor 
point, all the results discussed in the following 
qualitatively remain same for the simpler case of $\lambda=0$.
\begin{figure}
\begin{center}
\epsfig{file=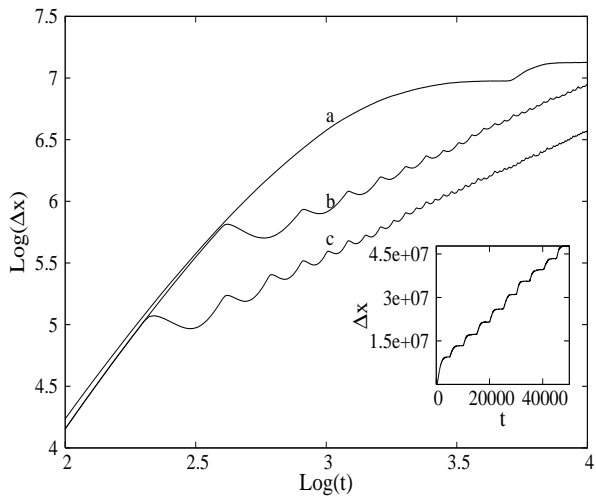, angle=-90, width=8.0cm,totalheight=6.5cm}
\caption{The position dispersion $\Delta x(t)$ for $\tau=5000$ (a),
400 (b), and 200 (c) are plotted. The Inset shows curve (a) 
extended to 10 half periods.}
\end{center}
\end{figure}

The particles exhibit coherent motion in the potential 
$V(x)=-\sin(x)-xF_0$, in the intermediate times roughly in the range 
$[\tau_1 (\approx 2\times 10^3)<t<\tau_2 (\approx 3\times 10^4)]$, for 
$F_0=0.2$. Therefore, we choose the ZMSW field $F(t)$ 
of amplitude $F_0=0.2$ and, in most cases, half period 
$\tau=5000$ which is well within the range $[\tau_1,\tau_2]$. Naturally, 
in the first half period $(0<t\leq\tau)$ the motion is same as in 
the CT case. In all cases, we take the initial ($t=0$) particle 
position distribution as $\delta(x-\frac{\pi}{2})$ and Maxwell 
velocity distribution corresponding to $T=0.4$. Note that 
after every half period $\tau$ the periodic potential gets tilted in the 
reversed direction as a result of field reversal.

Fig.1, curve (a) and its extended plot in the Inset show that by 
applying the field, $F(t)$, with $\tau=5000$ a 
repetitive sequence of trains of coherent motion, with characteristic 
constant $\Delta x(t)$ is obtained. These bursts of coherent 
motion are quite robust. Each burst of coherent motion is preceded by 
a length of dispersive particle motion. As a result $\Delta x(t)$ grows, 
in discrete steps, with time, as the number $n$ of half
periods increases. 
 
The generation of coherent motion continues for many ($n>>8$) half
periods ($\tau=5000$) of $F(t)$, Inset of Fig.1. Thus, as an important
consequence, the 
cumulative duration ($\approx n(\tau-\tau_1)$) of coherent motion when 
driven by $F(t)$ is made much larger than the
duration, $\tau_2-\tau_1$, achievable in the CT case.

In the Fig.1 are also plotted $\Delta x$ corresponding to the
half periods $\tau=$400 (curve (b)), and 200 (curve (c)). The important 
feature to be noticed in the figure is that, for 
$\tau=400, 200<\tau_K,\tau_1$, $\Delta x$ dips immediately after the 
field is reversed before it rises. This behaviour of dispersion dipping 
and subsequent rise becomes most pronounced at a small but intermediate 
$\tau$. It continues for many periods of $F(t)$.  

\begin{figure}
\begin{center}
\epsfig{file=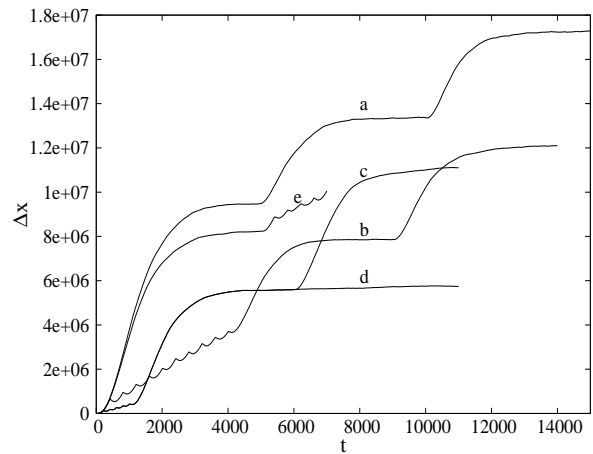, angle=-90, width=8.0cm,totalheight=6cm}
\caption{$\Delta x(t)$ for three $\tau=5000$ (a) ten $\tau=400$
and two $\tau=5000$ (b), five $\tau=200$ and two $\tau=5000$ (c),
five $\tau=200$ and one $\tau=10000$ (d), and one $\tau=5000$ and
five $\tau=400$ (e) half periods are plotted.}
\end{center}
\end{figure}

Fig.2, summarizes the main results of this work. It shows many 
interesting effects of taking few initial half periods of $F(t)$ 
of smaller duration, $\tau=400$ (curve (b)), and $\tau=200$ (curves 
(c-d)) and then making its later half periods $\tau=5000$ or larger 
(e.g., 10000, curve (d)). The curves (b-c) show 
that, in this case too, during the later half periods $\tau=5000$ 
of $F(t)$, a similar sequence of trains of coherent motion, 
as in Fig.1, Inset (or curve (a)), can be obtained. It also allows 
to postpone (curve (b)) the appearance of coherent motion beyond 
$t=\tau_1$. Moreover, it is possible to obtain coherent motion 
with lower constant dispersion (curves (c,d)) than in the CT case
(i.e., lower than the constant $\Delta x$ in the first half period 
in the curve (a)) too. Curve (e) shows that the curves (b-d) can be 
repeated, using the same customized procedure, many times over again.

The results of Figs.1 and 2 can be understood by analyzing the time 
evolution of velocity distribution, $P(v)$. $P(v)$ at various phases 
of $F(t)$ at $t=5\tau$, and at $t=\tau+15.6$ for $\tau=5000$ and at 
$t=2\tau$, and $t=2\tau+16$ for $\tau=1000$ are shown in Fig.3. $P(v)$ 
invariably assumes almost a Gaussian form of same width and centred at 
a fixed $v\approx\pm\frac{F_0}{\gamma_0}$ at $t=n\tau$ for $\tau=5000$.
The Inset, showing the mean velocity $\bar{v}(t)$ and velocity 
dispersion $\Delta v(t)$, supports this observation. However,
at $t=n\times 1000$, $P(v)$ has two peaks one centred at 
$v\approx\pm\frac{F_0}{\gamma_0}$ and the other at $v=0$. 

The $P(v)$ peak at 
$v=0$ shows that at $t=n\tau$, for $\tau<\tau_1$, some particles are 
left behind in the locked state in one or some other wells. The 
smaller is $\tau$ more prominent this latter peak is left at $t=\tau$. 
These locked particles try to gravitate to their respective well 
bottoms and, being slower, even succeed in shrinking the position 
distribution $P(x)$ more effectively than the much faster particles 
in the running state on the front. Thus, for smaller $\tau<\tau_1$, 
the bimodal nature of $P(v)$ helps appreciable $\Delta x$ dipping 
immediately after field reversal, Fig.1.
 
At $t=n\tau+t_0$, $15<t_0<16$, $\bar{v}$ becomes zero for all $\tau$, 
and $n$. (It turns out, $2\tau_0\approx\frac{m}{\gamma_0}$.) The 
corresponding $P(v)$ are also shown in Fig.3. For $\tau=5000$, 
$P(v)$ at $t=n\tau+t_0$ is like a sum of two "Gaussians" centred 
on either side of $v=0$. But for $\tau=1000$, the surviving $P(v)$ peak
at $v=0$ at $t=n\tau$, contributes an additional one centred at 
$v=0$, making $\bar{v}=0$ at the same delay time $\tau_0$. 

Due to the tilt direction reversal the particles, in the
running state, are forced to reverse their direction of motion and 
hence each of them necessarily go through zero velocity at least once 
momentarily. Thus, the entire system passes approximately through 
a "thermal" state at $t=n\tau+\tau_0$. Hence, after every field 
reversal at $t=n\tau$ the particles begin their subsequent journey 
in the reversed direction with almost the same delayed initial 
condition (at $t=n\tau+\tau_0$) of thermalized $P(v)$. Therefore, 
$\Delta x$ are expected to behave similarly after every $n\tau$.
 
As discussed above, it is just the reversal of field which leads 
$P(v)$ to the required form at $t=n\tau+t_0$, irrespective of the 
form of $P(v)$ at $t=n\tau$ for any value of $\tau>>\tau_0\approx16$. 
It shows that in order to obtain coherent motion neither an exact 
initial Gaussian velocity distribution is necessary nor all the 
particles are required to be intially confined sharply to a single 
well bottom of the periodic potential. Also, a mere switching the 
field alternately {\it on} ($F_0\neq0$) for duration $\tau$ and {\it off}
($F_0=0$) for the same duration $\tau$, fails to yield 
results like Figs.1 and 2. This indicates that a reversal of the field 
direction is essential because this alone ensures a "thermalized" 
$P(v)$.

\begin{figure}
\begin{center}
\epsfig{file=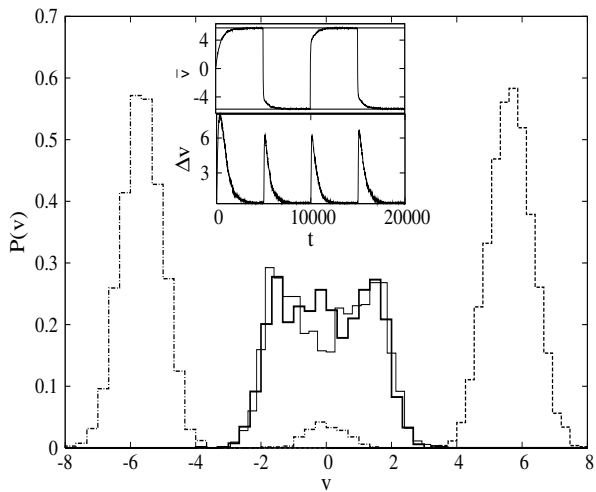, angle=-90, width=8.0cm,totalheight=6.5cm}
\end{center}
\caption{$P(v)$ at $t=25000$ (right peak)
and at 25015.6 (middle thin line) for
$\tau=5000$, and at $t=2000$ (bimodal), at 2016 
(middle bold line) for $\tau=1000$. $\bar{v}(t)$, and $\Delta v(t)$ 
for $\tau=5000$ are plotted ($\bar{v}=\pm F_0/\gamma_0$
lines drawn) in the Inset. }
\end{figure}

\begin{table*}
\caption[Table:]{The systems considered are: (1) Ag: Ag on AgI lattice, (2)
Mm: a macromolecule along a polymer, and (3) JJ: Josephson junction.
The symbols have their usual meaning. The RCSJ-model JJ equation
equivalent to Eq.(2.1) is\cite{falco}:\\
$\frac{\hbar C}{2e}\frac{d^2\theta}{dt^2}+\frac{\hbar}{2e}G(1+\lambda\cos(
\theta+\phi))\frac{d\theta}{dt}+I_1\sin(\theta)
=I(t)+\sqrt{2TG(1+\lambda\cos(\theta+\phi))}\xi(t)$.
(Add $-\frac{\pi}{2}$ to $\theta$ for exact correspondence.)}

\begin{tabular}{|l|c|c|c|c|c|c|c|c|c|c|} \hline
 & $m$(Kg) & $V_0$(eV) & $T$(K) & $k$(m$^{-1}$) 
& $\omega_0$(s$^{-1}$) & $\gamma_0$(Kg s$^{-1}$) 
& $\frac{\gamma_0}{m}$(s$^{-1}$) & $F_0$(N)
 & $\tau$(s) & $\bar{v}=\frac{F_0}{\gamma_0}$(ms$^{-1}$) \\ \cline{2-11}
$Ag$ & $1.79\times 10^{-25}$ & $0.15$ & $348$ 
& $0.5\pi\times 10^{10}$ & $4.07\times 10^{12}$ & $2.55\times 10^{-14}$ 
& $1.42\times 10^{11}$ & $5.03\times 10^{-10}$ & $1.23\times 10^{-9}$ 
& $1.48\times 10^3$\\ 
Mm & $3.32\times 10^{-22}$ & $0.13$ & $300$ & $2.5\pi\times 10^8$
 & $6.20\times 10^9$ & $7.21\times 10^{-14}$ 
& $2.17\times 10^8$ & $3.25\times 10^{-13}$
 & $8.06\times 10^{-7}$ & $4.51\times 10$ \\ \hline
 & $C$ & $I_1$ &$T$ & $\frac{2e}{\hbar}$
 & $\omega_P$ & $G$ & $\omega=\frac{G}{C}$
 & $I_0$ & $\tau$ & $\bar{V}=\frac{I_0}{G}$\\ \cline{2-11}

JJ & $0.5$ & $ 10^{-9}$ & $9.53$ & $3.038\times10^{15}$
 & $2.46\times 10^9$ & $4.31\times 10^{-5}$ & $8.63\times 10^7$
 & $2.0\times 10^{-8}$ & $2.03\times 10^{-6}$ & $4.64\times 10^{-4}$ \\
 & pF & Amp & mK & V$^{-1}$s$^{-1}$ & s$^{-1}$ & Ohm$^{-1}$
 & s$^{-1}$ & Amp & s & V\\ \hline
\end{tabular}
\end{table*}

It must also be noted that during the CT case the average 
particle displacement is large, $\approx\frac{\tau_2 F_0}{\gamma_0}$ 
by the end of its coherent motion whereas in the ZMSW 
case it is zero, for $\lambda=0$, and small and finite for 
$\lambda\neq 0$ and $\phi\neq0,\pi$ after any large time 
$t=2n\tau$. This is an added practical advantage over the CT case
for, in the ZMSW case, most of the particles on the average 
remain confined to a finite region of space despite periodically 
showing coherence of motion for a long time.

The dimensionless values of parameters used (e.g., $\gamma_0=.035,
T=.4, \tau=5000$, and $F_0=.2$) and other derived
quantities when restored to their usual units are presented in
Table 1 for three illustrative cases: (i) The motion 
of an Ag ion in AgI crystal\cite{hayana},
(ii) the motion of a macromolecule (kinesin) along a polymer fibre 
(microtubule)\cite{bier}, and (iii) diffusion of Cooper pairs across a 
Josephson junction\cite{falco}. Notice that 
$\omega(=\frac{\gamma_0}{m})<<\omega_0$, in the particle motion case, and 
$\omega<<\omega_P$, the Josephson plasma frequency 
($=\frac{2eI_1}{\hbar C}$), showing that the systems considered are, 
indeed, underdamped. The last column of the Table gives the magnitude of 
mean velocity (mean voltage) attained during the
coherent state when the initial value of the drive field $F(t)$
 ($I(t)$) is fixed either at $+|F_0|$ ($+|I_0|$) or with their sign
reversed and not an equal mixture of both. Also, during the
half period $\tau$ the particles move to an average distance (the product 
of quanties in the last two columns) of the 
order of a micron ($\mu$) which will get retraced in the next half period. This 
gives a rough idea of the sample size one would need to take in a 
ZMSW case. Also, $\frac{2\pi}{\tau}<<\omega_0 (\omega_P)$ 
(by about two orders of magnitude).

The calculated average velocities $\bar{v}$ and velocity dispersions 
$\Delta v$ are plotted, in the Inset of Fig.3.
An equal mixture of $F(t=0)=\pm|F_0|$ makes $\bar{v}$ close 
to zero during coherent motion. It is exactly zero for 
$\lambda=0$ at all $t$ and hence even in the limit $t\rightarrow\infty$ 
$\bar{v}$ remains zero. However, for $\lambda\neq 0$ and 
$\phi\neq 0,\pi$ a nonzero finite mean (steady state) velocity is 
obtained\cite{wanda,shantu,wanda1} earlier. The contribution of coherent 
particle motion being insignificant, the dispersive motion alone
contributes to this {\it ratchet} current of particles.
\begin{figure}
\begin{center}
\epsfig{file=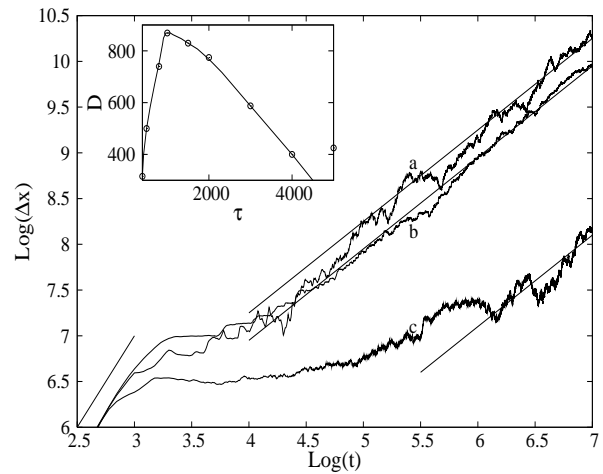, angle=-90, width=8.0cm,totalheight=6.2cm}
\caption{$\Delta x(t)$ for $\tau=$ 1000 (a), 5000 (b), and for the 
case of CT $F_0=0.2$ (c) averaged over, respectively, 20, 60, and 18
ensembles are plotted. Lines of slope 1 are fitted to the curves.
The short line at the lower left corner indicates $\Delta x\sim 
t^{\alpha}, \alpha\approx 2$.
The inset shows variation of $D$ with $\tau$.}
\end{center}
\end{figure}
  
The diffusion constant, $D$, defined as 
\( \lim_{t\rightarrow\infty}\Delta x(t)=2Dt\), is hard 
to calculate for a constant tilt $F_0$\cite{linde}. The asymptotic 
limit barely reaches even by $t=10^7$. However, for ZMSW 
this limit is readily reached by $t=10^7$, Fig.4. It 
may be noted that for each curve in Figs.1 and 2 we have averaged over 
2000 realizations but in Fig.4 we could average over number of 
realizations ranging only between 18 and 60. The nature of $\Delta x(t)$ 
so clear in Fig.1 appears less convincing in Fig.4. 
Therefore, it is hard to conclude that
the same nature of $\Delta x$, as in Fig.1, will continue till the
asymptotic time regime. However, from the "thermalized" $P(v)$ 
argument given earlier, there is a fair likelyhood that the nature of 
$\Delta x(t)$ shown in Fig.1 will
extend to a large number of half periods of $F(t)$ provided a large
number of particles are considered for averaging. 

The nature of
$\Delta x(t)$ shown by the curves in Figs.1, 2 is 
in no finite region close to $\Delta x(t)\sim t$. It remains, therefore, 
open to explain why a large number of repeatedly same diverse combinations 
of dispersions such as ones ranging from $\Delta x(t)\sim t^2$ to 
$\Delta x(t)\sim t^0$, when averaged over a large number of realizations, 
yields the same nature of dispersion, 
\( lim_{t\rightarrow\infty}\Delta x(t)\sim t \), for all $\tau$,
Fig.4.

$D(\tau)$, plotted in the Inset of Fig.4, are rough estimates as the 
averagings are done only over a small number of realizations. However, 
the overall qualitative trend of $D(\tau)$ remains valid. 
$D(\tau)$ shows a peak around $\tau=\tau_1$. For $\tau>\tau_1$, the 
closer $\tau$ is to 
$\tau_1$ smaller is the constant $\Delta x$ region and hence larger 
is the fraction of sharply rising $\Delta x$ region. Naturally total 
$\Delta x$ will be larger as $\tau\rightarrow\tau_1$. However, the nature
of $D(\tau)$ as $\tau\rightarrow\tau_2$ is not clear from the available
data. In the range $\tau<\tau_1$, $\Delta x$ rises only after
an appreciable dipping, Fig.1. Therefore, the initial rise of 
$\Delta x$ is 
slower as $\tau$ is decreased from $\tau_1$ resulting in a smaller 
total $\Delta x(t)$ as $t\rightarrow\infty$ and hence smaller $D$.

From the {\it rms} spread ($\sqrt{\Delta x}$) point of view the
advantage of ZMSW $F(t)$ over CT, except in cases like curve (d) 
in Fig.2, quickly evaporates as $t$ increases.  
Whereas (for an Ag particle) at $t=10\tau=12.3\times 10^{-9}$s for 
the CT (ZMSW) case, the mean displacement $\bar{x}$ is 
$1.8\mu$ ($\approx 0$) and 
$\sqrt{\Delta x}=.14\mu$ ($.40\mu$), at $t=2.46\times 10^{-6}$s the
corresponding numbers are $\bar{x}=3.64$mm ($\approx0$) and
$\sqrt{\Delta x}=.77\mu$ ($6.08\mu$). Perhaps in the ZMSW case, 
the particles left behind during a $\tau$ get pushed father away 
during the next $\tau$, make the $\sqrt{\Delta x(t)}$ increase 
faster as $t$ increases.
 
Coherent motion is observed only in the negative slope region of 
$D(\tau)$. However, for this same system it is shown in 
Ref.\cite{shantu} that ratchet current is maximum for a value of 
$\tau \simeq 500$, i.e., in the rising $D(\tau)$ region and becomes 
significantly small for larger $\tau\geq\tau_1$ and almost zero 
at $\tau=5000$. The peak of the $D(\tau)$, thus, roughly divides 
$\tau$ into two regions: (i) small $\tau$ giving ratchet current, 
and (ii) larger $\tau$ showing coherent motion.

To summarise, the dispersionless (coherent) motion discovered earlier, to 
occur for a brief but finite duration in the intermediate time regime, on 
a CT sinusoidal potential, was extended to the case of 
periodically reversing constant tilts. We have shown the possibility of 
obtaining coherent particle motion interspersed by dispersive motion over 
many periods of an external ZMSW field. The cumulative duration
of coherent motion can, thus, be extended to a substantial fraction of the 
total journey time, of course, at a cost of making the system several
times more diffusive.

MCM acknowledges support from the Abdus Salam ICTP, Trieste, where the 
paper was finalized.


\begin{thebibliography}{34}
\bibitem{risken}H. Risken, \emph{The Fokker-Planck Equation}, (Springer-Verlag, 
Berlin), 1996.

\bibitem{reim} P. Reimann, Phys. Rep. {\bf 361}, 57 (2002);
R.P. Feynman, R.B. Leighton, and M. Sands, {\it{The Feynman Lectures
in Physics}} (Publishers, year), Vol. 1, Chap. 46.

\bibitem{falco} C.M. Falco, Am. J. Phys. {\bf 44}, 733 (1976);
A. Barone, and G. Paterno, {\it Physics and Applications of the Josephson Effect},
John Wiley, New York, 1982.

\bibitem{fulde} P. Fulde, L. Pieternero, W.R. Schneider, and S. Str\"{a}ssler,
Phys. Rev. Lett. {\bf 35}, 1776 (1975).

\bibitem{lacasta} A.M. Lacasta, J.M. Sancho, A.H. Romero, I.M. Sokolov,
and K. Lindenberg, Phys. Rev. {\bf 70}, 051104 (2004).

\bibitem{linde} K. Lindenberg, J.M. Sancho, A.M. Lacasta, and I.M. Sokolov, Phys. Rev.
Lett. {\bf 98}, 020602 (2007).

\bibitem{wahn} G. Wahnstr\"{o}m, Surf. Sci. {\bf 159}, 311 (1985).

\bibitem{lambda} A.M. Jayannavar, and M.C. Mahato,
Pramana-J. Phys. {\bf 45} 369 (1995);
M.C. Mahato, T.P. Pareek, and A.M. Jayannavar,
Int. J. Mod. Phys. B{\bf 10}, 3857 (1996).

\bibitem{volmer} H. Risken, and H.D. Volmer, Z. Physik B {\bf 33}, 297 (1979).

\bibitem{wanda} W.L. Reenbohn, S. Saikia, R. Roy, and M.C. Mahato, Pramana J. Phys.
{\bf 71}, 297 (2008).


\bibitem{wanda1} W.L. Reenbohn, and M.C. Mahato, J. Stat. Mech: Theor. Exp. P03011
(2009).

\bibitem{shantu} S. Saikia, and M.C. Mahato, J. Phys.: Condens. Matter
{\bf 21}, 175409 (2009).

\bibitem{nume} W.H. Press, S.A. Teukolsky, W.T. Vetterling, and B.P. Flannery,
{\it Numerical Recipes (in Fortran): the Art of Scientific Computing}, Cambridge
University Press, Cambridge, 1992;
M.C. Mahato, and S.R. Shenoy, J. Stat. Phys. {\bf 73}, 123 (1993).

\bibitem{hayana} M. Hayana, A. Hatate, and M. Oguni, J. Phys.: Condens. Matter
{\bf 15}, 3867 (2003).

\bibitem{bier} R.D. Astumian, and M. Bier, Phys. Rev. Lett. {\bf 72}, 1766 (1994).

\end{thebibliography}
\end{document}